\documentclass[aps,prb,twocolumn,superscriptaddress,floatfix]{revtex4}
\usepackage{times}
\usepackage{graphicx}
\usepackage{epsfig}
\usepackage{amsmath}
\usepackage{amssymb}
\usepackage[usenames, dvipsnames]{color}
\usepackage[version=3]{mhchem}
\usepackage{rotating}

\begin{document}

\title{Local structure study of the orbital order/disorder transition in LaMnO$_3$}

\author{Peter M. M. Thygesen}
\affiliation{Department of Chemistry, University of Oxford, South Parks Road, Oxford OX1 3QR, U.K.}

\author{Callum A. Young}
\affiliation{Department of Chemistry, University of Oxford, South Parks Road, Oxford OX1 3QR, U.K.}

\author{Edward O. R. Beake}
\affiliation{Department of Chemistry, University of Oxford, South Parks Road, Oxford OX1 3QR, U.K.}
\affiliation{School of Physics and Astronomy, Queen Mary University of London,\\ Mile End Road, London E1 4NS, U.K.}

\author{Fabio Denis Romero}
\affiliation{Department of Chemistry, University of Oxford, South Parks Road, Oxford OX1 3QR, U.K.}

\author{Leigh D. Connor}
\affiliation{Diamond Light Source, Chilton, Oxfordshire, OX11 0DE, U.K.}

\author{Thomas E. Proffen}
\affiliation{Spallation Neutron Source, Oak Ridge National Laboratory, Oak Ridge, Tennessee 37831, U.S.A.}

\author{Anthony E. Phillips}
\affiliation{School of Physics and Astronomy, Queen Mary University of London,\\ Mile End Road, London E1 4NS, U.K.}

\author{Matthew G. Tucker}
\affiliation{Diamond Light Source, Chilton, Oxfordshire, OX11 0DE, U.K.}
\affiliation{Spallation Neutron Source, Oak Ridge National Laboratory, Oak Ridge, Tennessee 37831, U.S.A.}
\affiliation{ISIS Facility, Rutherford Appleton Laboratory,\\ Harwell Campus, Didcot, Oxfordshire OX11 0QX, U.K.}

\author{Michael A. Hayward}
\affiliation{Department of Chemistry, University of Oxford, South Parks Road, Oxford OX1 3QR, U.K.}

\author{David A. Keen}
\affiliation{ISIS Facility, Rutherford Appleton Laboratory,\\ Harwell Campus, Didcot, Oxfordshire OX11 0QX, U.K.}

\author{Andrew L. Goodwin}
\email{andrew.goodwin@chem.ox.ac.uk}
\affiliation{Department of Chemistry, University of Oxford, South Parks Road, Oxford OX1 3QR, U.K.}

\date{\today}
\begin{abstract}
We use a combination of neutron and X-ray total scattering measurements together with pair distribution function (PDF) analysis to characterise the variation in local structure across the orbital order--disorder transition in LaMnO$_3$. Our experimental data are inconsistent with a conventional order--disorder description of the transition, and reflect instead the existence of a discontinuous change in local structure between ordered and disordered states. Within the orbital-ordered regime, the neutron and X-ray PDFs are best described by a local structure model with the same local orbital arrangements as those observed in the average (long-range) crystal structure. We show that a variety of meaningfully-different local orbital arrangement models can give fits of comparable quality to the experimental PDFs collected within the disordered regime; nevertheless, our data show a subtle but consistent preference for the anisotropic Potts model proposed in \emph{Phys Rev.\ B} {\bf 79}, 174106 (2009). The key implications of this model are electronic and magnetic isotropy together with the loss of local inversion symmetry at the Mn site. We conclude with a critical assessment of the interpretation of PDF measurements when characterising local symmetry breaking in functional materials.

\end{abstract}


\maketitle


\section{Introduction}

Orbital degrees of freedom play a key role in the physics of colossal magnetoresistance (CMR),\cite{Millis_1996,Millis_1998,Rao_1996,Tokura_2000} frustrated magnetism,\cite{Broholm_2012,Yoshida_2012} ferroelectricity,\cite{Senn_2016,Stroppa_2011} spin glass formation,\cite{Thygesen_2017} and magnetoelectric coupling.\cite{Radaelli_1997} In all cases, it is the existence and nature of correlations between local orbital states that gives rise to the relevant phenomena of interest. Long-range orbital order can assume many forms, but its nature is usually evident crystallographically \emph{via} coupling between orbital occupancy/orientation and bond strain. This coupling results in long-range symmetry breaking, such as occurs in the cooperative Jahn-Teller (JT) state of KCuF$_3$.\cite{Suemune_1961} In favourable cases, resonant X-ray scattering also provides direct experimental evidence of long-range orbital order.\cite{Tokura_2004} By contrast, the microscopic nature of orbital-disordered states---as implicated in the phenomenology of CMR\cite{Ramirez_1997}---is notoriously difficult to determine experimentally. Certain probes (\emph{e.g.}\ EXAFS, PDF, NMR) remain sensitive to the presence of local orbital--strain coupling in the absence of long-range orbital order but their sensitivity to \emph{correlations} between orbital orientations in such states is either negligible or poorly understood.

Of the many systems known to exhibit orbital disorder, few can be more important than LaMnO$_3$. The parent of the La{$_{1-x}$}A{$_{x}$}MnO{$_3$} (A = Ca, Sr, Ba) families of CMR manganites,\cite{Rao_1996, Helmholt_1993,Millis_1996,Wollan_1955} LaMnO$_3$ has long assumed a special position amongst functional oxides.\cite{Jonker_1950,Goodenough_1955,Mathur_1997} Its orthorhombic--pseudocubic transition at $T_{\textrm{JT}}$ = 750\,K is widely viewed as the canonical orbital order/disorder transition.\cite{Goodenough_1999,Carvajal_1998,Billinge_2005} The same transition precedes CMR itself within the doped manganites\cite{Tokura_1994,Van_aken_2003, Billinge_2007,Lynn_2007} and is implicated more generally in charge and orbital ordering in a variety of other functional condensed phases.\cite{De_teresa_1997,Adams_2000,Radaelli_1995} 

Despite this importance, there remains no clear consensus regarding the microscopic nature of orbital disorder in LaMnO$_3$ itself. Zhou and Goodenough initially interpreted high-temperature resistivity and thermoelectric measurements in terms of dynamic cooperative JT distortions.\cite{Goodenough_1999} In this picture, orbital-driven distortions are similar in ordered and disordered states---the key difference is a transition from fixed to fluctuating collective orbital orientations. A conceptually similar model was proposed in the neutron PDF study of Qiu \emph{et al.}, where the orbital disordered phase was interpreted in terms of nm-sized domains with local (now static) orbital arrangements identical to those in the ordered state.\cite{Billinge_2005} These two models are consistent with conventional order--disorder descriptions: \emph{i.e.}, ordered and disordered states share a common local structure but temporal and/or configurational averaging leads to a finite correlation length in the disordered regime.\cite{Halcrow_2013} By contrast, Ahmed and Gehring reproduced the order parameter behaviour measured in resonant X-ray scattering experiments\cite{Endoh_1998} using the so-called anisotropic Potts model, which describes a transition to a meaningfully-different local arrangement of JT distortions at high temperatures.\cite{Ahmed_2009}


Motivated by the success of combined X-ray/neutron PDF measurements in characterising local orbital order in systems such as Y$_2$Mo$_2$O$_7$,\cite{Thygesen_2017} we sought to establish whether a similar approach might shed new light on the nature of orbital disorder in high-temperature LaMnO$_3$. The scattering contrast between X-ray and neutron measurements in principle heightens experimental sensitivity to different pairwise contributions to the PDF, and so is particularly useful in high-symmetry structures such as pyrochlores and perovskites.\cite{Reimers_1988} In this paper, we report a series of high-real-space-resolution X-ray and neutron PDF data collected across the LaMnO$_3$ orbital order--disorder transition. We show first the unsurprising result that these data unambiguously identify the nature of orbital order at temperatures below $T_{\textrm{JT}}$. We then report the discovery of a discontinuous change in the PDF at $T_{\textrm{JT}}$, overlooked in earlier PDF studies and ostensibly inconsistent with a conventional order--disorder transition. Within the high-temperature orbital-disordered regime, we find that the PDF data are in fact remarkably poor at discriminating models representing a variety of different locally-correlated orbital arrangements. We argue this insensitivity arises because the variation in pairwise correlations for different models become commensurate with the magnitude of thermal motion at these elevated temperatures. We cautiously identify a subtle but persistent preference for the anisotropic Potts model proposed in Ref.~\citenum{Ahmed_2009}. This assignment is consistent with the discontinuous change in local structure at $T_{\textrm{JT}}$ evident in our newly-obtained data. Our results suggest that short-range orbital correlations in the disordered phases of doped (CMR) manganites may be subtly but meaningfully different to those observed in the ordered regime.

Our paper is arranged as follows. We begin with a review of the current experiment-driven understanding of the orbital order/disorder transition in LaMnO$_3$, together with a structural description of the different correlated orbital arrangement models considered in our subsequent analysis. In section~\ref{methods} we summarise the experimental and analytical methods we have used, including a description of the key local structure models investigated in our study. We proceed to present our X-ray and neutron total scattering measurements, together with the results of conventional Rietveld analysis (average structure) and direct interrogation of the experimental PDFs (local structure). We then explore the ability of various orbital arrangement models to account for the PDF data we have measured within both orbital ordered and disordered regimes. Having established the difficulty of unambiguous interpretation of the disordered state, we consider the potential role of single-crystal diffuse scattering measurements in future studies. Our paper concludes with a short discussion of the implications of our study for CMR science, on the one hand, and for PDF studies of functional materials in general, on the other hand.

\section{Orbital order in lanthanum manganite}

\begin{figure}
\begin{center}
\includegraphics{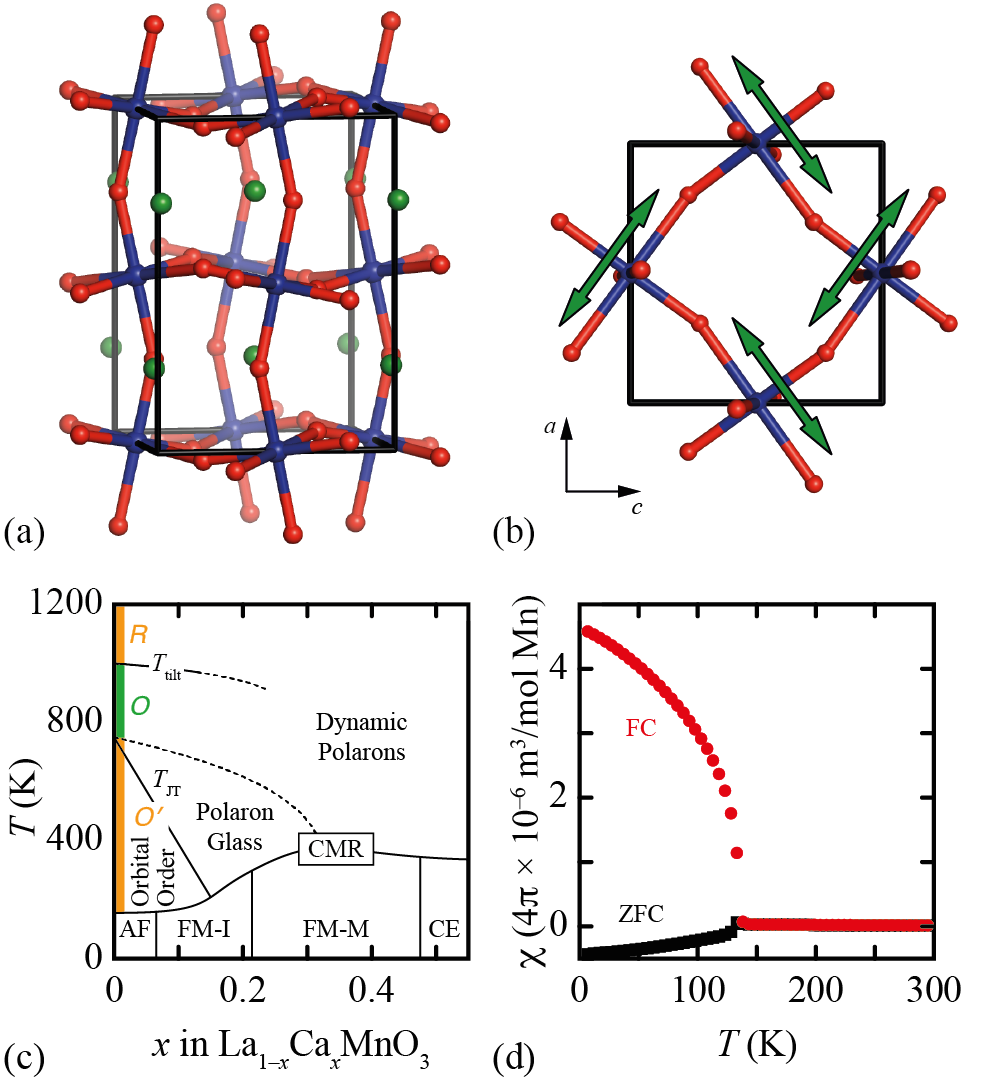}
\end{center}
\caption{\label{fig1} Structure and phase behaviour of LaMnO$_3$. (a) The ambient ($O^\prime$) structure consists of a framework of corner sharing MnO$_6$ octahedra (Mn atoms blue; O atoms red), with La atoms (green) occupying 12-coordinate extra framework sites. The presence of octahedral tilts confers orthorhombic $Pnma$ symmetry; in this representation the $b$ axis is vertical and the $c$ axis approximately horizontal. (b) Axial JT distortions are arranged with the same periodicity as the octahedral tilts. Long Mn--O bonds (indicated by arrows) are approximately confined within the $(a,c)$ planes; this particular arrangement results in a small difference between $a$ and $c$ lattice . (c) On heating LaMnO$_3$ to $T_{\textrm{JT}} = 750$\,K there is an isosymmetric structural transition to the orbital disorder ($O$) phase, with the same octahedral tilts as in the $O^\prime$ phase but with no discernible JT distortion. On hole doping, the value of $T_{\textrm{JT}}$ decreases such that CMR emerges in La$_{1-x}$Ca$_x$MnO{$_3$} from the disordered phase. A discontinuous octahedral tilt transition occurs at 1010\,K to give a rhombohedral ($R$) phase that persists until decomposition. (d) Experimental d.c. magnetic susceptibility $\chi$ for LaMnO$_3$ for the sample used in our study. The N\'{e}el temperature ($T_{\textrm N}$) is observed as a divergence in the field-cooled (FC) and zero-field-cooled (ZFC) traces at 140\,K.}
\end{figure}

The issue of orbital order in LaMnO{$_3$} arises fundamentally from the degeneracy of the Mn\textsuperscript{3+} \textit{t}\textsuperscript{3}\textit{e}\textsuperscript{1} d-electron configuration.\cite{Goodenough_1999} This degeneracy is lifted by a JT distortion of the [MnO{$_6$}] coordination environment, which acts to couple Mn--O bond displacements with orbital occupancies, and hence structural and electronic degrees of freedom.\cite{Goodenough_1998} Crystallographic measurements of LaMnO{$_3$} performed under ambient conditions indicate that these distortions (and hence orbital occupancies) are ordered throughout the crystal lattice, with a periodicity that coincides with that imposed by the octahedral tilt system also present.\cite{Carvajal_1998} This periodicity accounts for the orthorhombic $Pnma$ symmetry of the orbital ordered phase [Fig.~\ref{fig1}(a)]. There is a single Mn environment (Wyckoff site $4b$; site symmetry $\bar1$) and three symmetry-inequivalent pairs of Mn--O bonds: ``long'' (2.1\,\AA), ``medium'' (2.0\,\AA), and ``short'' (1.9\,\AA). The long and short bonds alternate within the $(a,c)$ plane [Fig.~\ref{fig1}(b)] to give an arrangement sometimes referred to as $C$-type orbital order.\cite{Matsumoto_1970} In this case, the specific octahedral tilts present mean that the projection of the long bonds is slightly greater along $a$ than $c$ and so the difference in these two lattice parameters is is actually related to the presence of orbital order.

This particular type of orbital order is known to have a direct effect on the transport and magnetic properties of ambient LaMnO{$_3$}:\cite{Goodenough_1999,Jonker_1956} the material is an insulator with strongly anisotropic magnetic interactions, as evidenced by a Curie-Weiss constant $\theta=52$\,K that represents a compromise between ferromagnetism within the $(a,c)$ plane and antiferromagnetic interactions between adjacent $(a,c)$ planes (\emph{i.e.}, along $b$).

\subsection*{Orbital order/disorder transition}\label{orderdisorder}

On heating from room temperature, LaMnO$_3$ undergoes two structural phase transitions [Fig.~\ref{fig1}(c)]. The first occurs at 750\,K and is isosymmetric---\emph{i.e.}, there is no change in space group symmetry.\cite{Carvajal_1998} This is the orbital order/disorder transition at the heart of this study. The second transition, at 1010\,K, is to a rhombohedral phase with a different set of tilts\cite{Carvajal_1998} but which also supports orbital disorder.\cite{Billinge_2005} Drawing on the approach taken in Ref.~\citenum{Goodenough_1955} we use the labels $O^\prime$, $O$, and $R$ to represent, respectively, the orthorhombic orbital ordered, orthorhombic orbital disordered, and rhombohedral orbital disordered phases.

Though isosymmetric, the $O^\prime/O$ transition at $T_{\textrm{JT}}=750$\,K has a number of clear experimental signatures. First, there is a convergence of the reduced unit cell parameters, such that the $O$ phase is close to being metrically cubic (hence its ``pseudocubic'' label).\cite{Carvajal_1998,Chatterji_2003} The variation in lattice parameters is discontinuous at $T_{\textrm{JT}}$, identifying the transition as first-order as required for isosymmetric transitions.\cite{Christy_1995} Second, both neutron and X-ray diffraction show a convergence of the crystallographic Mn--O bond lengths from the three distinct values of the ambient phase to a single effective value [$d(\textrm{Mn--O})\simeq2.02$\,\AA]  in the $O$ phase.\cite{Carvajal_1998,Chatterji_2003} So in this average-structure sense the JT distortion appears to vanish on heating through the $O^\prime/O$ transition. Third, there is a volume discontinuity at $T_{\textrm{JT}}$, with the disordered $O$ phase \emph{ca} 0.4\% denser than the ordered $O^\prime$ phase.\cite{Chatterji_2003} Fourth, both conductivity and magnetic behaviour change at the transition: resistivity falls by two orders of magnitude, and the Weiss constant switches to a value {$\theta$} = 177\,K that is consistent with isotropic ferromagnetic interactions within the $O$ phase.\cite{Goodenough_1999,Jonker_1956} Fifth, $^{17}$O NMR measurements reveal a transition to non-polarised $e_g$ orbital occupation at $T_{\textrm{JT}}$, indicating the population of MHz-frequency electronic fluctuations.\cite{Trokiner_2013} Sixth, there is an anomaly in the specific heat capacity, corresponding to a transition entropy $\Delta S=0.52(2)$\,J\,K$^{-1}$\,mol$^{-1}$.\cite{Sanchez_2003} And, seventh, the transition is accompanied by small shifts and increased broadening in the excitation spectrum as measured using either inelastic neutron scattering\cite{Wdowik_2012} or Raman spectroscopy.\cite{Iliev_1998}

The primary experimental evidence for the persistence of JT distortions within $O$-phase LaMnO$_3$ comes from techniques capable of probing structural correlations over distances of 1--10\,$\mathrm{\AA}$. The neutron PDF study of Ref.~\citenum{Billinge_2005} confirmed clearly the absence of any meaningful variation in local Mn--O bond lengths across the \emph{O}\textsuperscript{$\prime$}/\emph{O} transition. A similar conclusion was drawn from X-ray absorption (EXAFS/XANES) studies.\cite{Oseroff_2001,Sanchez_2003,Souza_2004} While there seems to be little disagreement as to the persistence of a JT distortion at high temperatures, it was noted first in Ref.~\citenum{Sanchez_2003} and more convincingly in Ref.~\citenum{Souza_2004} that the Mn K-edge EXAFS signal varied more strongly across the $O^\prime/O$ transition than for any temperature range within either phase. This is a point to which we will return later in our own study. Nevertheless, taken collectively, these experimental observations clearly identify the $O$ phase of LaMnO$_3$ as an orbital-disorder phase: each MnO$_6$ octahedron retains a JT distortion but these distortions must propagate without any long-range periodicity.

\subsection*{Orbital correlations in the $O$ phase}

Determining the nature of orbital orientation correlations in the absence of long-range orbital order is a problem of significant difficulty.\cite{Greedan_2009,Thygesen_2017,Billinge_2007,Keimer_2000,Keren_2001} Total scattering is one of the few experimental techniques with simultaneous sensitivity to both average and local structure,\cite{Billinge_2004,Young_2011} and so it is no accident that the two studies (of which we are aware) to attempt data-driven refinement of microscopic models of the orbital-disordered state are based on neutron total scattering measurements.\cite{Sartbaeva_2007,Billinge_2005} The first of these studies employed a ``real-space Rietveld'' (or ``small-box''; Ref.~\citenum{Egami_2003}) approach, in which the neutron PDF was fitted using a structural model based on the $O^\prime$ arrangement of orbital orientations.\cite{Billinge_2005} For data collected within the $O$ regime, convincing fits with stable JT distortions could only be obtained for refinements constrained to distances $0<r<10$\,\AA. Hence, the model developed in this study effectively describes the $O$-phase as an incoherent array of nm-sized domains, each with local orbital orientations as in the $O^\prime$ phase.

The second PDF study\cite{Sartbaeva_2007} made use of the same data set, but employed a custom ``big-box'' modelling approach\cite{Egami_2003,McGreevy_1988} based on a combination of geometric modelling and simulated annealing.\cite{Sartbaeva_2007} The key result was an atomistic configuration of LaMnO$_3$ that reproduced the experimental PDF while also preserving---within a predefined tolerance---the local geometry of all JT-distorted MnO$_6$ coordination polyhedra. In this configuration there was some evidence for locally-correlated orbital order of the antiferrodistortive $C$-type; however, the magnitudes of the relevant correlation functions were 50 times smaller than in the ordered phase itself.

\begin{figure}
\begin{center}
\includegraphics{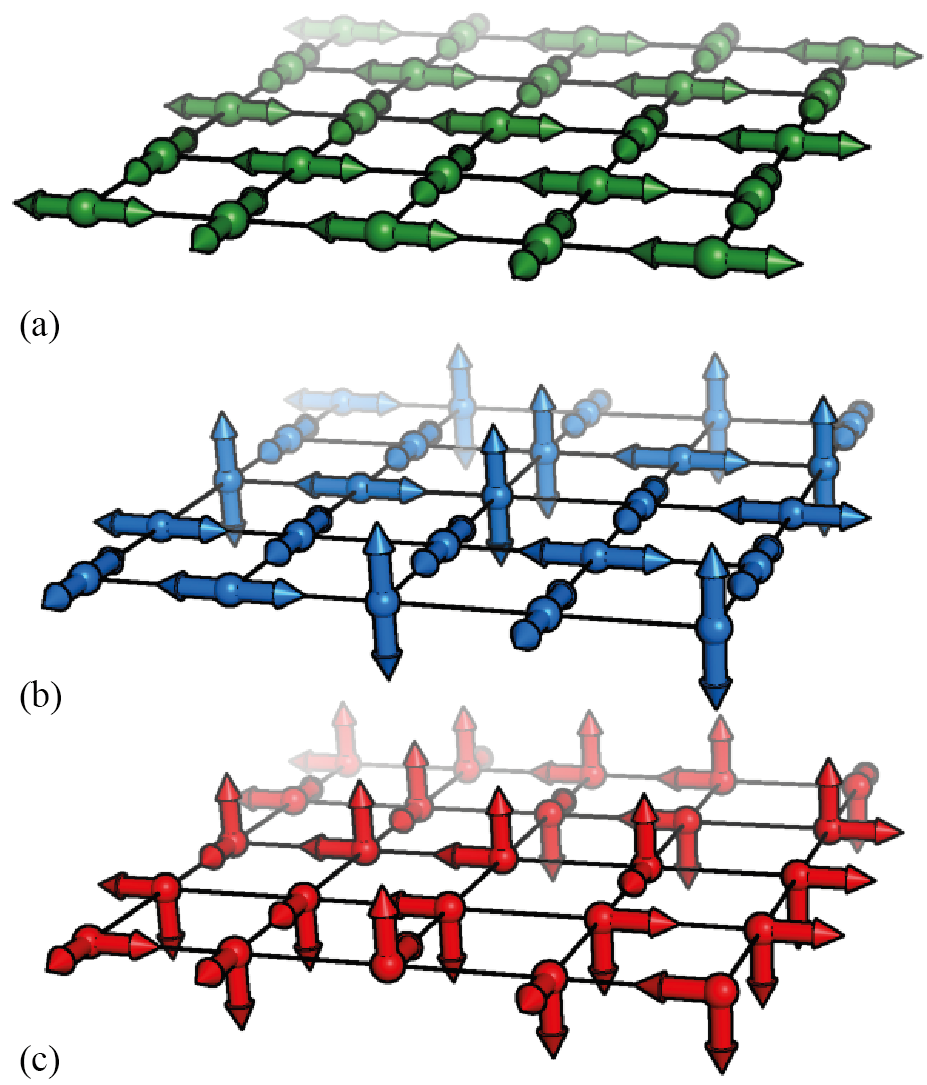}
\end{center}
\caption{\label{fig2} Simplified orbital orientations in the (a) $C$-type, (b) 3SP-type, and (c) $L$-type orbital models, shown here for a single planar section of the Mn sublattice (spheres). Arrows represent the orientation of long Mn--O bonds.}
\end{figure}

In hindsight, it is possible that the configurations obtained in this PDF study were actually realisations of the correlated disordered state of the anisotropic three-state Potts (3SP) model, as described in Ref.~\citenum{Ahmed_2009} and summarised here for completeness. The 3SP model is a statistical model in which each MnO$_6$ octahedron is free to adopt one of three possible states, corresponding to the three possible axes along which opposing pairs of long Mn--O bonds might orient. Neighbouring octahedra interact via two terms governing coupling of JT orientations, chosen so as to ensure $C$-type orbital order as the system ground state. On heating, the model undergoes a first-order phase transition to a disordered state in which antiferrodistortive coupling is (initially) strictly preserved but long-range orbital order is lost. Whereas the ground state samples only two of the three possible Potts states, the disordered state samples all three; this distinction means that the disordered phase supports local orbital orientation correlations that do not occur in the ordered regime. A visual comparison of the two states is given in Fig.~\ref{fig2}(a,b).

In anticipation of this study we had carried out our own ``big box'' modelling of the PDF data of Ref.~\citenum{Billinge_2005} (see SI for summary). Our approach had been to use the reverse Monte Carlo (RMC) method as implemented in RMCProfile.\cite{Tucker_2007} Whereas the study of Ref.~\citenum{Sartbaeva_2007} had constrained MnO$_6$ geometries in terms of the JT distortion found in the low-temperature $O^\prime$ phase, our own modelling employed only data-based closest-approach and ``distance-window'' constraints that allow much greater geometric flexibility.\cite{Tucker_2007,Goodwin_2005}  The unexpected but reproducible result of this analysis was a structural model in which the two long Mn--O bonds in each MnO$_6$ octahedron formed an angle of $\sim90^\circ$ rather than $\sim180^\circ$. Now adjacent to one another, the long Mn--O bonds form an ``L'' shape, and so we refer to this state as $L$-type orbital order [Fig.~\ref{fig2}(c)]. One effect of this bond arrangement is an off-centering of the Mn$^{3+}$ cation towards a single edge of its [MnO$_6$] coordination environment. Because there are 12 such edges, this arrangement would carry with it a much greater configurational degeneracy than either $C$- or 3SP-type orbital states. Our RMC configurations suggest not all such states are equally likely: instead a definitive ``ice-rules-like'' constraint\cite{Pauling_1935} emerged such that no two long Mn--O bonds were ever found to coincide at a single O atom. Even this local constraint leaves accessible a large configurational landscape associated with the $L$-type orbital order, as evidenced by the geometrically-related ``C2C'' procrystalline arrangement described in Ref.~\citenum{Overy_2016}. Though unexpected for LaMnO$_3$, this model is actually related to the ordered JT distortion known to occur in some Mn$^{3+}$ brownmillerites,\cite{Assey_2010,Wright_2002,Palmer_2006} and so could not be dismissed out of hand. 

One might expect the three candidate models for orbital orientation correlations in $O$-phase LaMnO$_3$ shown in Fig.~\ref{fig2} to be distinguishable in terms of their ability to account for experimental neutron and X-ray total scattering data. We proceed to assess precisely this point. Anticipating our results, we will come to show that the three models---as physically different as they are---give rise to remarkably similar fits to data as a result of the large magnitude of thermal motion for $T>T_{\textrm{JT}}$.

\section{Methods}\label{methods}
\subsection*{Sample preparation and characterisation}
A polycrystalline sample of stoichiometric LaMnO{$_3$} (6.5\,g) was prepared using the citrate gel method. Stoichiometric quantities of polycrystalline La$_2$O$_3$ and MnO$_2$ were dissolved in a minimum of 6\,M nitric acid. Two molar equivalents of citric acid and 5\,mL of ethylene glycol were added and the solutions heated with constant stirring.\cite{Hayward_2009} The resulting gel was dried and ground to fine powder, placed in a crucible and heated in air to 1000\,$^{\circ}$C for 10\,h. The powder was re-ground, pressed into pellets and fired at 1350\,$^{\circ}$C under flowing oxygen for 40\,h. Finally, the pellets were re-ground once more, pressed, and fired at 1350\,$^{\circ}$C under flowing argon. Phase purity was confirmed by X-ray diffraction and magnetisation measurements [Fig.~\ref{fig1}(d)]. The N\'{e}el temperature ($T_{\textrm N}$) was confirmed to be 140\,K for our sample, which is in good agreement with previously-reported values.\cite{Ritter_1997}

\subsection*{Neutron total scattering}
Neutron total scattering data were collected using the GEM diffractometer at the ISIS pulsed neutron and muon source.\cite{Hannon_2005} The polycrystalline LaMnO$_3$ sample was loaded into a cylindrical vanadium can of 8\,mm diameter and 5.8\,cm height, and placed in a custom-designed furnace. Total scattering data were collected over the reciprocal-space range $0.7\leq Q\leq40$\,\AA$^{-1}$, corresponding to a real-space resolution of order $\Delta r\simeq3.791/Q_{\textrm{max}}=0.09$\,\AA. Data collection times were optimised for total scattering measurements; we performed a total of six such measurements at temperatures of 523, 653, 753, 823, 903, and 973\,K.

Following collection, the total scattering data were corrected and placed on an absolute scale using standard methods as implemented in the GUDRUN software.\cite{Soper_2011} In this process we took into account the effects of background scattering, absorption, multiple scattering, and beam intensity variations. As an additional precaution---and motivated by the significant phonon populations at the high temperatures at which our measurements were carried out---we explored the effects of successively including and omitting Placzek inelasticity corrections,\cite{Dove_2002} and found no meaningful sensitivity to these corrections in our data. The Bragg profile function was extracted from the scattering data collected by the detector banks centred on $2\theta=34^\circ, 62^\circ, 92^\circ,$ and $146^\circ$ and used as input for Rietveld analysis in the GSAS software.\cite{Toby_2001} The corrected total scattering data were also transformed to the pair distribution function (we used the ``$G^{\textrm{PDF}}(r)$'' normalisation as defined in Ref.~\citenum{Keen_2000}), which was then used as input for real-space Rietveld analysis in the PDFGui software.\cite{Farrow_2007}

\subsection*{X-ray total scattering}
Variable-temperature X-ray total scattering data were measured using the high-energy I12 beamline at the Diamond Light Source\cite{Drakopoulos_2015} (X-ray wavelength $\lambda=0.14577(1)$\,\AA). A small fraction of the same polycrystalline sample of LaMnO$_3$ used for neutron scattering measurements was finely ground and loaded into a 1\,mm quartz capillary. The capillary was mounted vertically on a rotatable goniometer, and the sample temperature adjusted using a hot air blower calibrated \emph{in situ} with a thermocouple. A beam size of 1.5\,mm $\times$ 1.5\,mm was selected, and a Thales Pixium RF4343 2D detector (CsI scintillator, 430\,mm $\times$ 430\,mm) mounted 474\,mm from the sample collected the data. Only the top half of the detector was used to avoid problems caused by pixel mismatch at the connection between the two halves. The beam was centred in the corner of this half to allow scattering to be detected to as high an angle as possible. X-ray total scattering data were collected at 10\,K intervals from 523\,K to 973\,K. For each temperature point, 600 one-second exposures were collected and then averaged. Background measurements were obtained using the scattering from an empty quartz capillary in an otherwise identical setup. Initial data processing was carried out using the Fit2D software.\cite{Hammersley_1996} A mask was used to remove dead pixels in the detector, and each image was integrated to give a one dimensional scattering pattern. The integrated diffraction pattern was used directly for Rietveld refinements of the average crystal structure, making use of the TOPAS refinement software.\cite{Coelho} For PDF refinements, the integrated X-ray scattering data were used as input for the suite of standard background corrections and data normalisation processes as performed using GudrunX.\cite{Hannon_1990} This process yielded the normalised X-ray PDF as defined in Ref.~\citenum{Keen_2000}. The usable maximum magnitude $Q_{\textrm{max}}$ of the X-ray total scattering function was 22.5\,\AA$^{-1}$.


For ease of representation we use the term $G(r)$ to refer to both neutron and X-ray PDFs in the particular normalisations outlined above.

\subsection*{Comparative PDF refinements}\label{models}
PDF refinements were carried out using the PDFGui software as described in Ref.~\citenum{Farrow_2007}. In our study we compare the quality of fits for three orbital-correlation models: $C$-type, 3SP-type, and $L$-type, as summarised in Fig.~\ref{fig2}. In order to allow statistically-meaningful direct comparison amongst fits, we identified a suitable small-box approximant for each model such that all three models shared a common number of structural degrees of freedom. This approximant approach is described fully in Ref.~\citenum{Thygesen_2017}, but we proceed to summarise the key points here. The concept is to fit the low-$r$ PDF data in terms of a small-box model derived from the known average structure by a specifically-chosen symmetry-lowering perturbation. While it is understood that this symmetry-breaking process can in general propagate in so many different ways as to give a large manifold of degenerate disordered states, the approximant represents the highest-symmetry arrangement of the particular perturbation of interest. In this way the number of refineable parameters is kept to a minimum. Hence one expects the quality of PDF fit to decrease as the $r$-range included in the fit increases. In our PDF fits, we compare the results for three combinations of PDF data: (i) $1.5\leq r\leq6$\,\AA\ neutron $G(r)$, (ii) $1.5\leq r\leq6$\,\AA\ neutron + X-ray $G(r)$, and (iii) $1.5\leq r\leq10$\,\AA\ neutron + X-ray $G(r)$. The PDFGui parameters $\delta_1$, $Q_{\textrm{damp}}$ (X-ray) and $Q_{\textrm{damp}}$ (neutron) were fixed at values of 1.7\,\AA, 0.0486\,\AA$^{-1}$, and 0.0343\,\AA$^{-1}$, respectively.



We proceed to describe the three specific approximant models used in our comparative PDF refinements. In each case, we identify the highest-symmetry perturbation of a common parent structure that captures the relevant local orbital correlations for a given model and then parameterise the orbital-driven JT distortion using a single, refineable, distortion parameter (we call this $\delta$). The common parent, or reference, structure was determined as follows. Conventional Rietveld refinement against X-ray and neutron Bragg diffraction data collected at each temperature gave an average structure model in the space group $Pnma$. For each temperature point, the corresponding octahedral tilt magnitude $\phi$ was then determined from the oxygen atom positions according to the approach outlined in Refs.~\citenum{Keeffe_1977, Carvajal_1998}. The reference structure for a given temperature is then also described by the space group $Pnma$---with precisely the same lattice parameters as determined by Rietveld refinement and with La and Mn atoms on their conventional sites ($4c$ and $4b$ Wyckoff positions, respectively)---but with the oxygen positions now determined solely by the octahedral tilt angle:\cite{Keeffe_1977}
\begin{eqnarray}
x_{\textrm{O1}}&=&\frac{\cos^2\phi-1}{2\cos^2\phi+4}\nonumber\\
y_{\textrm{O1}}&=&\frac{1}{4}\nonumber\\
z_{\textrm{O1}}&=&\frac{\sqrt{3}+\tan\phi}{\sqrt{12}}\nonumber\\
x_{\textrm{O2}}&=&\frac{2-\sqrt{3}\sin\phi\cos\phi+\cos^2\phi}{8+4\cos^2\phi}\nonumber\\
y_{\textrm{O2}}&=&-\frac{\tan\phi}{\sqrt{48}}\nonumber\\
z_{\textrm{O2}}&=&\frac{3\sqrt{3}+\tan\phi}{\sqrt{48}}\label{coords}
\end{eqnarray}
In this way the reference structure effectively represents the Rietveld-refined $Pnma$ structure with JT distortions removed.

\subsubsection*{$C$-type orbital correlations}

\begin{figure}
\begin{center}
\includegraphics{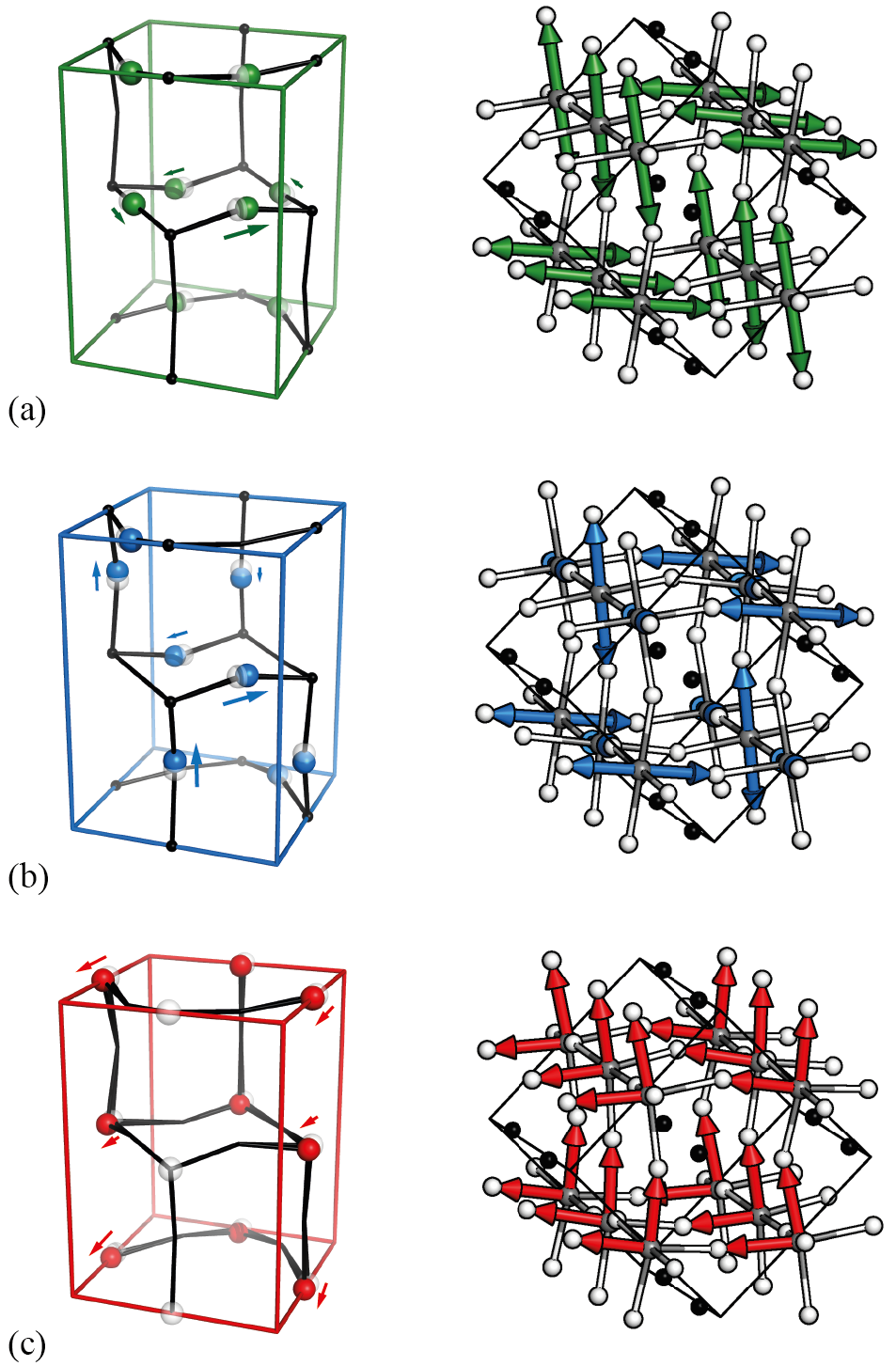}
\end{center}
\caption{\label{fig3} Representation of the small-box approximants used in our PDF study of orbital disordered LaMnO$_3$. (a) $C$-type order is generated by collective displacement of O2 atoms (green arrows, left), preserving the $Pnma$ symmetry of the undistorted parent structure. (b) 3SP-type order is generated by collective displacements of both O1 and (some) O2 atoms, as indicated by the blue arrows in the left-hand panel. This arrangement has $P2_1/n11$ symmetry (c) $L$-type order results from collective displacement of Mn atoms (red arrows, left) to give an approximant with $Pnm2_1$ symmetry. In each case, the corresponding arrangement of long Mn--O bonds is shown by the arrows in the right-hand panels.}
\end{figure}

For the small-box approximant used to model $C$-type orbital correlations, the relevant distortion parameter must give rise to the correlated JT distortion pattern observed in the (orbital ordered) $O^\prime$ phase as observed experimentally. This type of orbital order has the same space-group symmetry as that of the activated tilts, so this approximant shares the $Pnma$ symmetry of the reference structure. The distortion parameter $\delta$ is implemented in PDFGui as a variation in the O2 atom position, giving the local orbital correlation pattern shown in Fig.~\ref{fig3}(a). The relationship between atom positions and positional parameters in the reference and approximant models is given explicitly in Table~\ref{table1}.

PDFGui refinements using this approximant model make use of total of ten variable parameters (see Ref.~\citenum{Farrow_2007} for definitions): (i) $Q_{\textrm{broad}}$, (ii) an overall scale factor, (iii) a relative scale factor for X-ray \emph{vs} neutron data sets, (iv, v) the $x_{\textrm{La}}$ and $z_{\textrm{La}}$ positional parameters of the La atom site, (vi) the JT distortion parameter $\delta$, and (vii-x) isotropic displacement parameters for each of the La, Mn, O1, and O2 sites.

\subsubsection*{3SP-type orbital correlations}
For the 3SP-type orbital model we use the highest-symmetry subgroup of $Pnma$ which permits JT distortions in the out-of-plane ($b$-axis) direction. Distortion mode analysis using the ISOTROPY software suite\cite{Campell_2011} identified $P2_1/n11$ as the relevant subgroup, where we have used an unconventional setting of the $P2_1/c$ space group in order to facilitate direct comparison between reference and approximant structures. The monoclinic distortion angle $\alpha$ was fixed at $90^\circ$ for the same reason. So as to describe a 3SP-type collective JT distortion in terms of the single parameter $\delta$ it was necessary for this parameter to vary the positions of both O1-type and O2-type oxygen atoms. Our specific implementation is given in Table~\ref{table1}, and was chosen so as to ensure similar magnitudes of in-plane and out-of-plane JT distortions. The corresponding orbital correlation pattern is shown in Fig.~\ref{fig3}(b).  By design, the total number of variable parameters for this approximant is identical to that for the $C$-type model described above.

 \begin{table*}
\begin{center}
\caption {Relationship between atom positions in reference and approximant small-box models as described in the text. For each model, and for each atomic site, the relevant Wyckoff position is given together with the atom coordinates used as input in PDFGUi refinements. Whereas the values of the parameters $x_{\textrm{La}}$ and $z_{\textrm{La}}$ were taken directly from the results of our Rietveld refinements, the values $x_{\textrm{O1}},z_{\textrm{O1}},x_{\textrm{O2}},y_{\textrm{O2}},z_{\textrm{O2}}$ were determined using Eq.~\eqref{coords}. Only those parameters marked by an asterisk were allowed to vary during PDFGui refinements; variable parameters with the same symbol for a given model were constrained to assume identical values. All models share the same unit cell axes, and hence lattice parameters.}
\begin{tabular}{l|cc|cc|cc|cc}
\hline\hline
&\multicolumn{2}{c|}{Reference structure}&\multicolumn{2}{c|}{$C$-type}&\multicolumn{2}{c|}{3SP-type}&\multicolumn{2}{c}{$L$-type}\\
\hline
Space group and origin shift&\multicolumn{2}{c|}{$Pnma; [0,0,0]$}&\multicolumn{2}{c|}{$Pnma; [0,0,0]$}&\multicolumn{2}{c|}{$P2_1/n11; [0,0,0]$}&\multicolumn{2}{c}{$Pnm2_1; [\frac{1}{4},\frac{1}{4},0]$}\\
\hline\hline
La&$4c$&$(x_{\textrm{La}},\frac{1}{4},z_{\textrm{La}})$&$4c$&$(x_{\textrm{La}}^\ast,\frac{1}{4},z_{\textrm{La}}^\ast)$&$4e$&$(x_{\textrm{La}}^\ast,\frac{1}{4},z_{\textrm{La}}^\ast)$&$2a$&$(x_{\textrm{La}}^\ast,\frac{1}{4},z_{\textrm{La}}^\ast)$\\
& & & & & & &$2a$&$(-x_{\textrm{La}}^\ast,\frac{3}{4},-z_{\textrm{La}}^\ast)$\\
\hline
Mn&$4b$&$(0,0,\frac{1}{2})$&$4b$&$(0,0,\frac{1}{2})$&$2b$&$(0,0,\frac{1}{2})$&$4b$&$(0,0,\frac{1}{2}+\delta^\ast)$\\
& & & & &$2d$&$(\frac{1}{2},0,0)$& & \\
\hline
O1&$4c$&$(x_{\textrm{O1}},\frac{1}{4},z_{\textrm{O1}})$&$4c$&$(x_{\textrm{O1}},\frac{1}{4},z_{\textrm{O1}})$&$4e$&$(x_{\textrm{O1}},\frac{1}{4}-\delta^\ast,z_{\textrm{O1}})$&$2a$&$(x_{\textrm{O1}},\frac{1}{4},z_{\textrm{O1}})$\\
& & & & & & &$2a$&$(-x_{\textrm{O1}},\frac{3}{4},-z_{\textrm{O1}})$\\
\hline
O2&$8d$&$(x_{\textrm{O2}},y_{\textrm{O2}},z_{\textrm{O2}})$&$8d$&$(x_{\textrm{O2}}+\delta^\ast,y_{\textrm{O2}},z_{\textrm{O2}}+\delta^\ast)$&$4e$&$(x_{\textrm{O2}}+\delta^\ast,y_{\textrm{O2}},z_{\textrm{O2}}+\delta^\ast)$&$4b$&$(x_{\textrm{O2}},y_{\textrm{O2}},z_{\textrm{O2}})$\\
& & & & &$4e$&$(x_{\textrm{O2}}+\frac{1}{2},y_{\textrm{O2}},\frac{1}{2}-z_{\textrm{O2}})$&$4b$&$(-x_{\textrm{O2}},-y_{\textrm{O2}},-z_{\textrm{O2}})$\\
\hline\hline
\end{tabular}
\label{table1}
\end{center}
\end{table*}

\subsubsection*{$L$-type orbital correlations}

Our third and final approximant model corresponds to the alternate JT distortion that had emerged from our preliminary RMC analysis: \emph{i.e.}\ angles of \emph{ca} 90$^\circ$ between the two long Mn--O bonds at each Mn site. Distortion mode analysis using the ISOTROPY suite identified $Pnm2_1$ as a relevant subgroup of highest symmetry. (Again we use an unconventional setting of the space group $Pmn2_1$ to facilitate comparison amongst our various small-box models). In our preliminary RMC configurations, there was little evidence for distortion of the O$_6$ coordination octahedra around each Mn; rather the JT distortion was accommodated by displacement of the Mn atom away from the centre of each such octahedron. Consequently, the distortion parameter $\delta$ no longer influences the O1 or O2 atom positions in this approximant model, but instead displaces the Mn atoms towards an edge of the corresponding octahedral MnO$_6$ coordination environments. Our RMC refinements indicated a preference for locally-ferroelectric Mn displacement patterns; the $Pnm2_1$ approximant is the simplest model that captures these Mn correlations. This distortion is shown in Fig.~\ref{fig3}(c) and the implementation in PDFGui is given explicitly in Table~\ref{table1}. We note for completeness that the total number of free variables is again conserved for this model. More complex approximants in which Mn displacements included out-of-plane components are conceivable but would require a larger number of variables to be refined.

\subsection*{Single-crystal diffuse scattering calculations}
Single-crystal diffuse scattering patterns were generated from explicit atomistic realisations of the 3SP- and $L$-type orbital disorder states. In the former case it is the positions of apical O atoms that determines orbital orientations; in the latter case the key component involves Mn displacements. So that we might calculate diffuse scattering for large ($20\times20\times20$) supercells---and hence obtain smoothly continuous diffraction patterns---we excluded the ordered component from our atomistic configurations. The role of dynamic disorder was not taken into account. The calculations themselves were carried out using the SingleCrystal software.\cite{Palmer}

\section{Results and discussion}

\subsection*{Neutron and X-ray diffraction: The average structure}

\begin{figure}
\begin{center}
\includegraphics{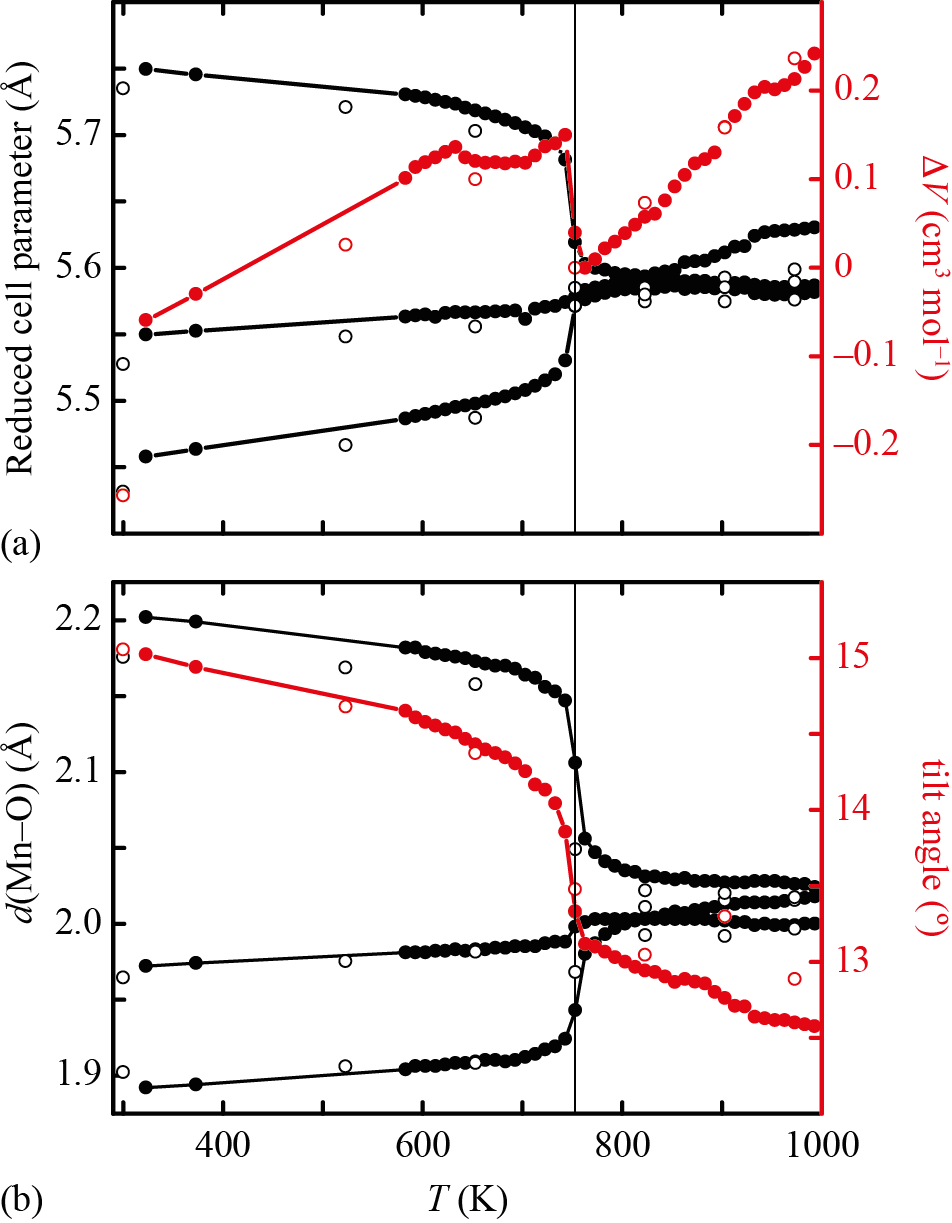}
\end{center}
\caption{\label{fig4} Temperature-dependent structural variation in LaMnO$_3$ as determined using Rietveld analysis of X-ray (filled symbols) and neutron (open symbols) total scattering data. (a) The unit cell metric becomes pseudocubic at $T_{\textrm{JT}}$ with a volume collapse of \emph{ca} 0.15\,cm$^{3}$\,mol$^{-1}$. (b) The three crystallographically-distinct Mn--O bond lengths also converge at $T_{\textrm{JT}}$; a modest variation in tilt angle also accompanies the transition.}
\end{figure}

Our starting point was to determine the temperature-dependent behaviour of the average structure of LaMnO$_3$ over the temperature range $300\leq T\leq1000$\,K, as reflected by the Bragg component of the neutron and X-ray total scattering functions. Using Rietveld refinement, we determined the variation in unit cell parameters with temperature; our results [Fig.~\ref{fig4}(a)] clearly identify the isosymmetric phase transition between $O^\prime$ and $O$ phases on heating above $T_{\textrm{JT}}\simeq750$\,K. We used this transition to cross-calibrate the sample temperatures of X-ray and neutron measurements. We find the transition to be accompanied by a volume collapse of $-0.1499$\,cm$^{3}$\,mol$^{-1}$ [Fig.~\ref{fig4}(a)], which is similar to the value of $-0.1535$\,cm$^{3}$\,mol$^{-1}$ noted in Ref.~\citenum{Chatterji_2003}. The reciprocal-space resolution of our measurements is marginally poorer than that used in the study of Ref.~\citenum{Chatterji_2003} (a consequence of optimising for real-space resolution)---this discrepancy is likely responsible for the small unphysical fluctuation in molar volume near $T_{\textrm{JT}}$. The most reliable structural models were obtained using Rietveld refinement of data collected at those temperatures for which we had access to both X-ray and neutron scattering measurements. The results of our refinements for the seven such temperature points spanning $O^\prime$ and $O$ phases are given in Table~\ref{table2}; these values are consistent with those reported in Refs.~\citenum{Carvajal_1998}. As anticipated from our discussion in Section~\ref{orderdisorder}, we find the key consequences of orbital disorder within the $O$ phase to be (i) the emergence of a pseudocubic lattice metric, (ii) the near equivalence of Mn--O bond lengths, and (iii) the increase in atomic displacement parameters.

\begin{table*}
\begin{center}
\caption {Structural parameters for LaMnO$_3$ as determined by Rietveld refinement against both neutron and X-ray Bragg diffraction patterns. Included are the two crystallographically-distinct Mn--O--Mn bond angles ($\theta_1,\theta_2$), from which the octahedral tilt parameter $\phi$ was determined according to the method outlined in Ref.~\citenum{Carvajal_1998}.}
\begin{tabular}{l|ccccccc}
\hline\hline
$T$/K&300&523&653&753&823&903&973\\
Phase&$O^\prime$&$O^\prime$&$O^\prime$&$O$&$O$&$O$&$O$\\
\hline
$a$/\AA & 5.73484(4) & 5.72072(4)& 5.70290(5) & 5.58482(13) & 5.57448(11) & 5.57446(7) & 5.57592(8)\\ 
$b$ /\AA & 7.68113(5) & 7.73080(7) & 7.75978(8) & 7.8791(3) & 7.8910(2) &  7.89888(18) & 7.90448(18)\\  
$c$/\AA & 5.52744(3) & 5.54801(4) & 5.55559(4) & 5.57204(18) &  5.58503(16) & 5.59239(9) & 5.59864(9)\\
\hline
$x_\textrm{La}$ & 0.04884(7) & 0.04397(8) & 0.04041(10) & 0.02522(17) & 0.02136(12) & 0.02026(14) &  0.01703(14)\\
$z_\textrm{La}$ & 0.00802(8) & 0.00683(10) & 0.00624(12) & 0.0041(5) &  0.0036(3) & 0.0023(5) & 0.0014(3)\\  
$x_\textrm{O1}$ & 0.98790(9) & 0.98869(12) & 0.98973(14) &  0.9905(3) & 0.9902(3) & 0.9921(3) & 0.9911(3)\\ 
$z_\textrm{O1}$ & 0.57395(9) & 0.57258(12) & 0.57183(13) &  0.5691(5) & 0.5692(3) & 0.5708(4) &  0.5675(3)\\    
$x_\textrm{O2}$ & 0.19346(7) & 0.19565(7) & 0.19789(8) & 0.2156(3) & 0.22025(19) & 0.2241(2) & 0.22411(18)\\ 
$y_\textrm{O2}$ & 0.96418(5) & 0.96210(6) & 0.96274(6) & 0.9619(2) & 0.96226(14) & 0.96384(19) & 0.96247(14)\\ 
$z_\textrm{O2}$ & 0.77436(8) & 0.77297(10) & 0.77200(12) & 0.7741(3) & 0.7758(2) & 0.7756(2) &  0.7730(2)\\ 
$U_\textrm{iso}$(La)/\AA$^2$ & 0.00520(12) & 0.00365(12) &  0.00420(13) & 0.00610(18) & 0.01381(19) & 0.01595(13) & 0.0162(2)\\ 
$U_\textrm{iso}$(Mn)/\AA$^2$ & 0.00227(19) &  0.00058(19) & 0.0013(2) & 0.0038(3) & 0.00874(19) & 0.01020(17) & 0.0111(2)\\ 
$U_\textrm{iso}$(O1)/\AA$^2$  & 0.00679(17) &  0.00647(18) & 0.0087(2) & 0.0152(5) & 0.0228(4) & 0.0266(5) & 0.0274(5)\\ 
$U_\textrm{iso}$(O2)/\AA$^2$ & 0.00603(12) & 0.00482(12) &  0.00649(14) & 0.0130(3) & 0.0210(3) & 0.0238(3) & 0.0253(4)\\
\hline
$\theta_1/^\circ$ & 155.63 & 156.17 & 156.51 & 157.67 & 157.62 & 156.95 & 158.16\\ 
$\theta_2/^\circ$ & 155.12 & 155.81 & 156.47 & 158.24 & 158.76 & 159.52 & 159.68\\ 
$\phi/^\circ$ & 15.06 & 14.68 & 14.37 & 13.48 & 13.33 & 13.30 & 12.89\\   
 \hline\hline
			\end{tabular}
		\label{table2}
	\end{center}
\end{table*}

Although the sensitivity of X-ray scattering data to O atom positions is markedly reduced when interrogated in the absence of accompanying neutron scattering data, we nevertheless obtained satisfactory Rietveld refinements at all temperature points probed in our X-ray study. In this way we obtained insight into crystallographic variations across the orbital order/disorder transition at a much finer temperature interval than was feasible from our neutron scattering measurements. In Figure~\ref{fig4}(b) we show the evolution of Mn--O bond lengths and octahedral tilt angles as a function of temperature across this transition. We find the expected discontinuous convergence of Mn--O bond lengths, together with a small variation in tilt angle at $T_{\textrm{JT}}$ that (to the best of our knowledge) has not been reported in previous studies.\cite{Carvajal_1998,Chatterji_2003} Our conclusion in these various respects is simply that our sample and the temperature dependence of its average structure are entirely consistent with previous observations of stoichiometric LaMnO$_3$.

\subsection*{Neutron and X-ray PDF: Direct analysis}

\begin{figure}
\begin{center}
\includegraphics{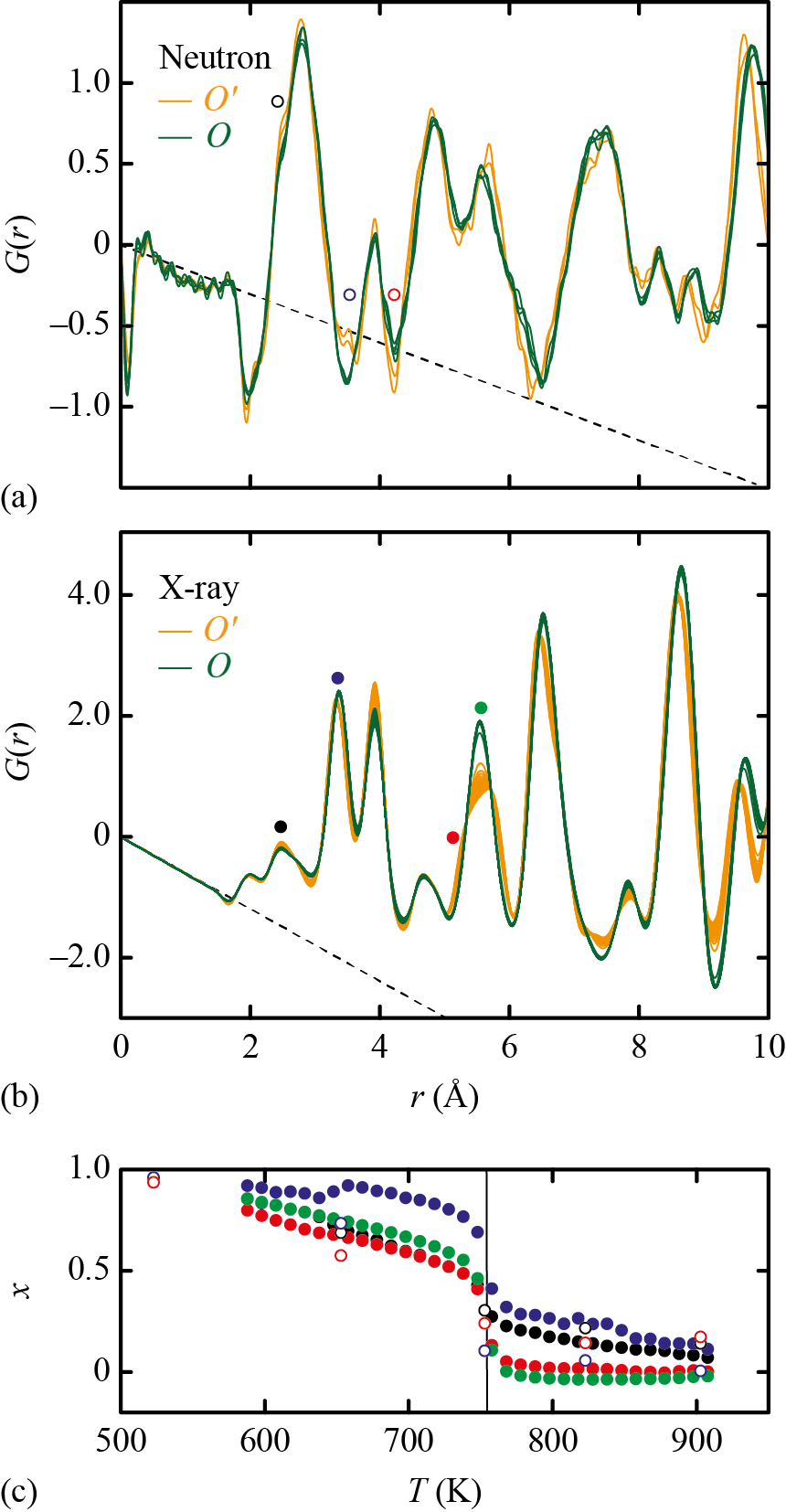}
\end{center}
\caption{\label{fig5} Temperature-dependent variation in (a) neutron and (b) X-ray PDFs across the $O^\prime$ (orange lines) / $O$ (green lines) transition. In both panels (a) and (b) the PDF baseline is indicated as a dashed black line. (c) The normalised temperature dependence of PDF intensities $x$ (see main text for derivation) at a variety of key interatomic separations clearly reveals a discontinuous change in local structure accompanying the orbital order--disorder transition. The relevant interatomic separations are indicated in panels (a) and (b) using open and filled circles (neutron and X-ray data, respectively); these symbols correspond to those used in panel (c) to represent the corresponding temperature dependence.}
\end{figure}

The temperature-dependent variation in neutron and X-ray $G(r)$ functions is shown in Fig.~\ref{fig5}. These data represent the key new experimental results of our study. We note in particular that the previous PDF studies of Refs.~\citenum{Sartbaeva_2007} do not include X-ray PDF measurements; moreover the neutron PDFs as reported do not allow for direct comparison at distances beyond the nearest Mn--O separation. While our X-ray PDF data show little sensitivity to pairwise correlations involving O atom positions, their temperature-dependent behaviour is nevertheless indirectly sensitive to cooperative JT phenomena \emph{via} coupling between O displacements and Mn and La atom positions. In our data, we find excellent consistency in PDFs obtained from independent measurements at different temperatures, with a high ratio of signal-to-noise. The persistence of JT distortions within the orbital-disordered $O$ phase is directly evident in the asymmetry of the nearest-neighbour Mn--O peak of the neutron PDF ($r\simeq2$\,\AA; note the peak is inverted as a consequence of the negative neutron scattering length of Mn). The reduced real-space resolution of our X-ray PDF data, together with the low X-ray scattering power of Mn and O atoms (relative to La) means that we cannot resolve the JT distortion directly from our X-ray PDF measurements. Nevertheless, in these respects our data are entirely consistent with the results of earlier PDF studies.\cite{Billinge_2005,Sartbaeva_2007}

What is particularly notable in our new PDF data is the clear signature of a discontinuous change in PDFs---both neutron and X-ray---across the $O^\prime/O$ orbital order/disorder transition that involves correlations at interatomic distances commensurate with the unit cell dimensions. This discontinuity is evident in the raw PDF data themselves [Fig.~\ref{fig5}(a,b)] but is even more clearly identified by considering the relative changes in $G(r)$ at specific values of $r<6$\,\AA. For a given separation of interest $r^\prime$, we used separate linear fits to the $G(r=r^\prime,T)$ function for temperatures $T<T_{\textrm{JT}}$ and $T>T_{\textrm{JT}}$ to identify expectation values $\tilde G_{O^\prime}$ and $\tilde G_{O}$ at 500 and 1000\,K (\emph{i.e.}, deep within the $O^\prime$ and $O$ regimes, respectively). Using these expectation values we form the local order parameter\cite{Knox_2014}
\begin{equation}
x(r^\prime,T)=\frac{G(r^\prime,T)-\tilde G_{O}}{\tilde G_{O^\prime}-\tilde G_{O}},
\end{equation}
which (by construction) tends to unity as $T\rightarrow500$\,K and to zero as $T\rightarrow1000$\,K. This process allows us to compare meaningfully the smoothness in variation of $G(r)$ with $T$ for different interatomic separations across both X-ray and neutron PDFs. We calculated $x(r^\prime,T)$ functions for seven representative pairwise separations $r^\prime$ in both neutron and X-ray PDFs; the results are shown in Fig.~\ref{fig5}(c) and clearly identify a discontinuous variation in $G(r)$ at $T_{\textrm{JT}}$. This behaviour is entirely consistent with the EXAFS and XANES studies of Refs.~\citenum{Sanchez_2003,Souza_2004}, which also measured a greater difference in signal across $T_{\textrm{JT}}$ than for comparable temperature intervals within either the $O^\prime$ or $O$ phases.

In conventional order/disorder transitions, such as that between the $\alpha$- and $\beta$- phases of SiO$_2$ cristobalite\cite{Tucker_2000} or the primitive and body-centered phases of Cu$_3$Au,\cite{Proffen_2000,Proffen_2002} the PDF is unchanged at the transition itself for short distances. This is because the local structure in both ordered and disordered states is essentially identical. At high temperatures, the local distortions present within the disordered state are not correlated over macroscopic length-scales, but the correlation length diverges on cooling towards the transition temperature. Hence such transitions are accompanied by discontinuities in the Bragg diffraction pattern (which is sensitive to long-range correlations), without any abrupt change in the accompanying PDF (sensitive to short-range correlations). It is only at larger separations (\emph{e.g.}\ the Si$\ldots$Si separation in cristobalite\cite{Tucker_2000}) that significant changes in the PDF are evident. Instead the behaviour we observe for LaMnO$_3$ at $T_{\textrm{JT}}$ reflects a local structure transition, as observed in \emph{e.g.} the metal--organic frameworks UiO-66 and NU-1000 during solvent removal,\cite{Chapman_2016} or indeed across the metal--insulator transition in VO$_2$ (Ref.~\citenum{Corr_2010}) or the doped manganites La$_x$Ca$_{1-x}$MnO$_3$.\cite{Billinge_2007} Consequently we can be confident that the orbital order/disorder transition in LaMnO$_3$ involves a discontinuous change in local structure, which in turn implies that orbital arrangements within the high-temperature $O$ phase may be meaningfully different to those in the ordered $O^\prime$ phase.\footnote{We note that we have tested the robustness of the behaviour shown in Fig.~\ref{fig5}(c) to variations in data processing. A particular concern is the role of inelastic scattering at elevated temperatures. Irrespective of the particular normalisation protocol we use for treating our X-ray or neutron total scattering data, we obtain essentially identical trends in the behaviour of $G(r)$ at $T_{\textrm{JT}}$.} In this respect, we anticipate that the $C$-type orbital correlation model should not provide a meaningful description of local orbital correlations in the $O$ phase. We comment that, given that isosymmetric transitions are intrinsically first-order,\cite{Christy_1995} it would be difficult to rationalise the bulk thermodynamic anomalies of LaMnO$_3$ at $T_{\textrm{JT}}$ in the absence of any discontinuity in local structure.



\subsection*{Neutron and X-ray PDF: PDFGui analysis}

In order to achieve a deeper understanding of the local orbital order in $O$-phase LaMnO$_3$, we proceeded to carry out a series of small-box PDF refinements using the PDFGui suite of programs.\cite{Farrow_2007} We consider the three orbital correlation models described in Section~\ref{models}, and interpret the contrasting ability of these models to account for both X-ray and neutron $G(r)$ functions at distances $\leq6$\,\AA. The importance of this particular 6\,\AA\ distance is twofold: first, it is commensurate with the unit cell dimensions, and so represents a length-scale that is large enough to be sensitive to differences in the various models we study, but also small enough that the use of approximants remains valid; and, second, it includes the various interatomic separations highlighted in Fig.~\ref{fig5}(c) for which we know $G(r,T)$ to be discontinuous at $T_{\textrm{JT}}$. For completeness we will also compare these results with fits to neutron data alone, and also with fits to combined neutron and X-ray PDFs carried out over the larger range of separations $1.5\leq r\leq10$\,\AA. Such comparisons will allow us insight into the robustness of any conclusions drawn with respect to the particular relative weighting of X-ray and neutron data, on the one hand, and value of $r_{\textrm{max}}$ over which fitting is carried out, on the other hand. 

\begin{figure}
\begin{center}
\includegraphics{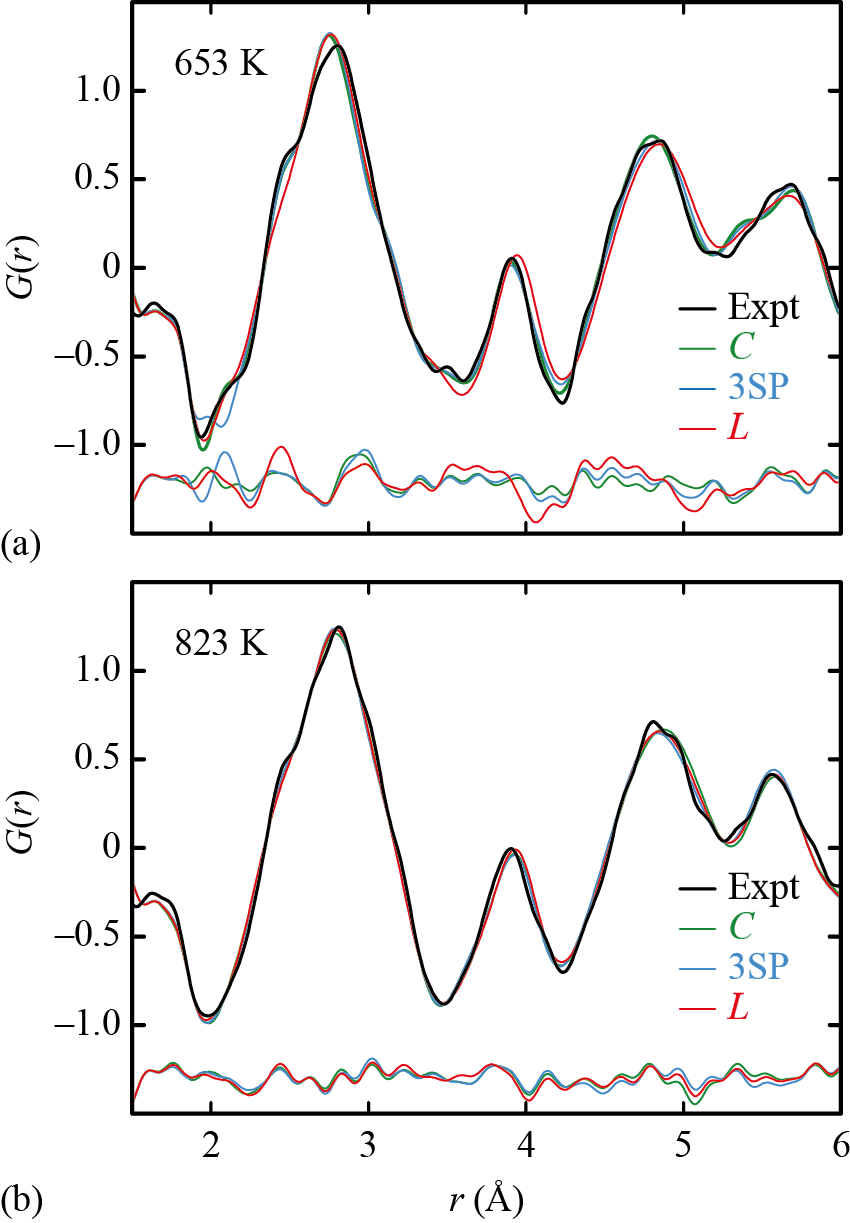}
\end{center}
\caption{\label{fig6} Neutron PDF fits for data collected within the (a) $O^\prime$ and (b) $O$ regimes. Experimental data are shown as black lines, and fits shown in green ($C$-type order), blue (3SP-type order) and red ($L$-type order). The corresponding difference functions (data $-$ fit) are shown in the relevant colour, shifted vertically by 1.25 units. The fits shown here were for PDFGui refinements involving simultaneous fits to X-ray PDFs (see SI).}
\end{figure}

Within the orbital-ordered $O^\prime$ regime the PDF data show a clear preference for the $C$-type orbital correlation model. We demonstrate this point in Fig.~\ref{fig6}(a), where we compare the quality of fits for all three models to the neutron $G(r)$ collected at 653\,K. Despite the meaningful differences amongst the three models, and the clear indication from conventional (average structure) interpretation of the diffraction pattern that LaMnO$_3$ exhibits $C$-type orbital order at this temperature, the differences in quality of fit for the three models are probably much smaller than intuition might have suggested. Nevertheless the $L$-type model clearly results in a significantly poorer fit throughout the whole $r$-region, and the 3SP-type model cannot capture the correct shape of the nearest-neighbour Mn--O peak while also fitting the remainder of the PDF. The refined parameters and qualities of fit are given explicitly in Table~\ref{table3}.
 
The magnitude of JT distortion is accounted for by the fitting parameter $\delta$, which is related to the difference between long and short Mn--O bond lengths. Simple geometric considerations give
\begin{equation}
\Delta d(\textrm{Mn--O})\simeq k|\delta|\sqrt{a^2+c^2},
\end{equation}
where the proportionality factor $k=2$ for the $C$- and 3SP-type models, and $k=1$ for the $L$-type model. For the $C$-type model fitted against neutron and X-ray PDF data collected at 653\,K (\emph{i.e.}\ within the orbital order regime) we obtain a JT distortion of $0.25$\,\AA, which is remarkably similar to the value obtained from Rietveld refinement against Bragg intensities [$\Delta d(\textrm{Mn--O})\simeq0.25$--0.3\,\AA; Fig.~\ref{fig4}(c)]. Despite their poorer fits, both 3SP- and $L$-type models give similar distortions, showing this result is robust to choice of modelling approach. These results are entirely consistent with earlier PDF studies.\cite{Billinge_2005,Sartbaeva_2007}

\begin{figure}
\begin{center}
\includegraphics{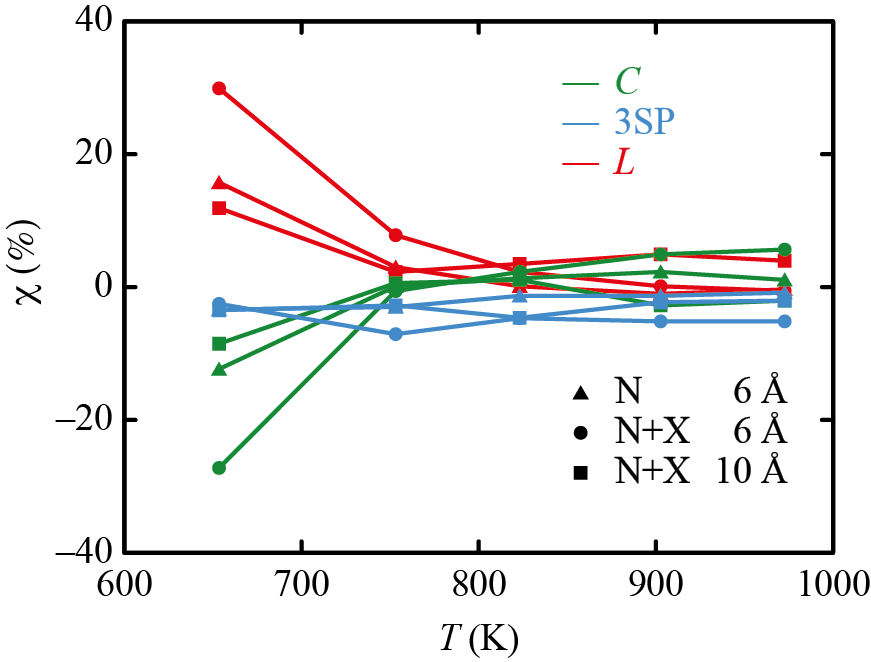}
\end{center}
\caption{\label{fig7} Relative goodness-of-fit $\chi$ for the various PDF fits described in the text. The three orbital arrangement models are given in the same colours as in Fig.~\ref{fig6}. The different symbols correspond to different fitting protocols: triangles indicate fits to neutron PDF data alone over the real-space range $1.5\leq r\leq6$\,\AA, circles to both neutron and X-ray data over the same range, and squares to both data sets but over the increased range $1.5\leq r\leq10$\,\AA. We expect the approximant approach to become less reliable at larger $r$-values, so the most reliable fits correspond to those shown using circles; details of the various refined parameters for these fits are given in Table~\ref{table3}.}

\end{figure}

Within the orbital disorder regime, the situation is substantially less clear-cut. In Fig.~\ref{fig6}(b) we compare the fits for our three models against the neutron $G(r)$ function collected at 823\,K---\emph{i.e.}, within the $O$ phase. These fits are remarkably similar and cannot be discriminated by eye. Numerically, we find the 3SP-type orbital correlation model describes the data best, but the corresponding $R_{\textrm{wp}}$ value is only marginally lower than that for either the $C$- or $L$-type models [Table~\ref{table3}]. Moreover, different fitting protocols---neutron only \emph{vs} neutron + X-ray, or $r_{\textrm{max}}=6$\,\AA\ \emph{vs} 10\,\AA--- result in variations in $R_{\textrm{wp}}$ that are of the same order of magnitude as the differences between candidate models. We quantify this point by comparing in Fig.~\ref{fig7} the relative quality of fit
\begin{equation}
\chi=\frac{R_{\textrm{wp}}}{\langle R_{\textrm{wp}}\rangle}-1
\end{equation}
for our three models and three different fitting regimes as a function of temperature. The value of $\chi$ reflects the extent to which a particular model fits more closely ($\chi<0$) or less closely ($\chi>0$) than the average fit obtained for a given temperature point. So at 653\,K, for example, the $C$-type orbital model gives the best fit irrespective of the particular fitting protocol adopted. For temperatures above $T_{\textrm{JT}}$, however, the best-fit model can depend on the fitting protocol, which means that it is not possible to identify unambiguously the orbital arrangement pattern within the $O$ phase from quality of PDF fit alone. We attribute this insensitivity to the large magnitude of thermal motion at temperatures for which the orbital-disordered state is observed. If we recalculate the PDFs for each model using the same parameters obtained in our PDFGui refinements but reduce the magnitude of thermal parameters to ambient-temperature values, then much clearer differences amongst the various fits emerge (see SI).

\begin{table*}
\begin{center}
\caption {Model parameters for LaMnO$_3$ as determined by PDFGui refinement against both neutron and X-ray PDFs. The highest-quality fit for each temperature point is highlighted in bold: we find the $C$-type approximant to best describe the orbital ordered state (as anticipated from the average structure), and the 3SP-type approximant to best describe the orbital disordered state, at all temperature points investigated.}
\begin{tabular}{lc|cccccccccccc}
\hline\hline
&$T$/K	&	$Q_{\textrm{broad}}$/\AA$^{-1}$	&	scale1	&	scale2	&	$x_{\textrm{La}}$	&	$z_{\textrm{La}}$	&	$\delta$	&	\textit{U}$_\mathrm{iso}$(La)/\AA$^2$	&	\textit{U}$_\mathrm{iso}$(Mn)/\AA$^2$	&	\textit{U}$_\mathrm{iso}$(O1)/\AA$^2$	&	\textit{U}$_\mathrm{iso}$(O2)/\AA$^2$	&	$R_{\textrm{wp}}$/\%\\
\hline
$C$	&	{\bf 653}	&	{\bf 0.19(12)}	&	{\bf 0.162(16)}	&	{\bf 0.21(6)}	&	{\bf 0.036(6)}	&	{\bf 0.0120(8)}	&	{\bf $-$0.016(5)}	&	{\bf 0.008(7)}	&	{\bf 0.013(15)}	&	{\bf 0.02(2)}	&	{\bf 0.025(12)}	&	{\bf 12.3875}	\\
	&	753	&	0.19(13)	&	0.159(18)	&	0.20(5)	&	0.022(9)	&	0.009(14)	&	$-$0.014(7)	&	0.008(9)	&	0.02(2)	&	0.03(5)	&	0.037(19)	&	10.7698	\\
	&	823	&	0.18(13)	&	0.160(19)	&	0.20(5)	&	0.019(11)	&	0.010(14)	&	$-$0.014(8)	&	0.009(10)	&	0.02(2)	&	0.04(7)	&	0.04(2)	&	10.9087	\\
	&	903	&	0.17(12)	&	0.159(19)	&	0.20(5)	&	0.017(12)	&	0.009(16)	&	$-$0.013(9)	&	0.010(10)	&	0.02(2)	&	0.05(9)	&	0.04(3)	&	10.8627	\\
	&	973	&	0.13(11)	&	0.16(2)	&	0.18(5)	&	0.012(17)	&	0.00(5)	&	0.013(13)	&	0.015(12)	&	0.03(2)	&	0.04(11)	&	0.06(5)	&	10.8156	\\ \hline
3SP	&	653	&	0.18(12)	&	0.161(16)	&	0.21(6)	&	$-$0.036(6)	&	0.012(8)	&	$-$0.014(5)	&	0.008(6)	&	0.017(18)	&	0.014(18)	&	0.028(15)	&	13.6527	\\
	&	{\bf 753}	&	{\bf 0.18(13)}	&	{\bf 0.159(18)}	&	{\bf 0.20(5)}	&	{\bf $-$0.022(9)}	&	{\bf 0.007(18)}	&	{\bf $-$0.014(7)}	&	{\bf 0.009(10)}	&	{\bf 0.02(2)}	&	{\bf 0.03(5)}	&	{\bf 0.04(2)}	&	{\bf 10.4497}	\\
	&	{\bf 823}	&	{\bf 0.15(12)}	&	{\bf 0.161(19)}	&	{\bf 0.20(5)}	&	{\bf $-$0.018(11)}	&	{\bf 0.01(3)}	&	{\bf 0.013(8)}	&	{\bf 0.011(12)}	&	{\bf 0.02(2)}	&	{\bf 0.03(7)}	&	{\bf 0.05(4)}	&	{\bf 10.6293}	\\
	&	{\bf 903}	&	{\bf 0.15(12)}	&	{\bf 0.161(19)}	&	{\bf 0.20(5)}	&	{\bf $-$0.016(13)}	&	{\bf 0.00(4)}	&	{\bf 0.015(10)}	&	{\bf 0.013(12)}	&	{\bf 0.03(2)}	&	{\bf 0.03(8)}	&	{\bf 0.05(5)}	&	{\bf 10.4781}	\\
	&	{\bf 973}	&	{\bf 0.13(12)}	&	{\bf 0.16(2)}	&	{\bf 0.18(5)}	&	{\bf $-$0.014(17)}	&	{\bf 0.00(6)}	&	{\bf 0.012(10)}	&	{\bf 0.015(12)}	&	{\bf 0.03(2)}	&	{\bf 0.04(5)}	&	{\bf 0.08(17)}	&	{\bf 10.6128}	\\
 \hline
$L$	&	653	&	0.16(12)	&	0.164(17)	&	0.21(5)	&	$-$0.034(8)	&	0.011(17)	&	0.033(12)	&	0.011(8)	&	0.013(19)	&	0.02(3)	&	0.04(2)	&	16.3559	\\
	&	753	&	0.16(10)	&	0.161(18)	&	0.20(5)	&	$-$0.020(12)	&	0.01(3)	&	0.027(17)	&	0.012(9)	&	0.02(3)	&	0.04(8)	&	0.04(3)	&	11.0794	\\
	&	823	&	0.15(10)	&	0.162(18)	&	0.20(5)	&	$-$0.017(14)	&	0.01(4)	&	0.028(19)	&	0.013(10)	&	0.02(4)	&	0.05(11)	&	0.05(4)	&	10.7938	\\
	&	903	&	0.15(10)	&	0.161(19)	&	0.20(5)	&	$-$0.016(15)	&	0.01(4)	&	0.03(2)	&	0.014(12)	&	0.02(4)	&	0.05(13)	&	0.05(5)	&	10.5081	\\
	&	973	&	0.14(10)	&	0.16(2)	&	0.18(5)	&	$-$0.013(18)	&	0.00(4)	&	0.03(2)	&	0.016(14)	&	0.02(5)	&	0.06(18)	&	0.05(6)	&	10.6526	\\
\hline\hline
			\end{tabular}
		\label{table3}
	\end{center}
\end{table*}

Nevertheless we can draw some general conclusions from the results shown in Fig.~\ref{fig7}. Throughout the orbital disorder regime, the 3SP-type arrangement nearly always gives the highest-quality fit-to-data, even if the relative $\chi$ values for the $C$- and $L$-type models depends on the particular fitting protocol used. Moreover, for what we consider to be the most meaningful choice of fitting parameters (neutron and X-ray data, with $r_{\textrm{max}}=6$\,\AA), the 3SP-type model consistently represents the data most closely. Consequently we cautiously suggest that the local orbital arrangements in orbital-disordered LaMnO$_3$ are more accurately described by the 3SP model of Ref.~\citenum{Ahmed_2009} than either the $C$-type arrangement of the ambient-temperature phase---\emph{i.e.}\ as proposed in the PDF study of Ref.~\citenum{Billinge_2005}---or the $L$-type arrangement that emerges from RMC analysis.

This conclusion is supported by two other observations. First, we might have already reasonably ruled out the $C$-type arrangement on the basis that we observe a discontinuous change in PDF at $T_{\textrm{JT}}$. Were the orbital arrangements within the high-temperature phase to share the same pattern as the low-temperature ordered phase, then there would be no reason for such a discontinuity. And, second, of the two models with different orbital arrangements to that of the ordered phase, it is the 3SP-type arrangement that contains the more conventional local JT distortion of MnO$_6$ octahedra. While the off-centered distortion of the $L$-type model is not without precedent in the structural chemistry of manganites,\cite{Assey_2010,Wright_2002,Palmer_2006} the 180$^\circ$ arrangement of long Mn--O bonds is unquestionably the more frequently observed JT distortion for octahedral Mn$^{3+}$.\cite{Goodenough_1998}

\subsection*{Single-crystal diffuse scattering}

In principle one might expect greater sensitivity in distinguishing these three models using single-crystal diffuse scattering measurements than PDFs, which are derived from the same scattering function but only after orientational averaging.\cite{Welberry_2016} We are not aware of any published single-crystal scattering patterns (either X-ray or neutron) for LaMnO$_3$ within the orbital disorder regime, and the collection of such data was beyond the scope of our own study. Nevertheless the qualitative form of structured diffuse scattering anticipated for different orbital disorder models is straightforwardly calculated and may be of use in future investigations. We proceed to present the results of two such calculations for the 3SP and $L$-type disorder models, noting that the $C$-type orbital disorder model would result only in a diffuse component centred on Bragg reflections of the parent orbital-ordered $Pnma$ phase.

\begin{figure}
\begin{center}
\includegraphics{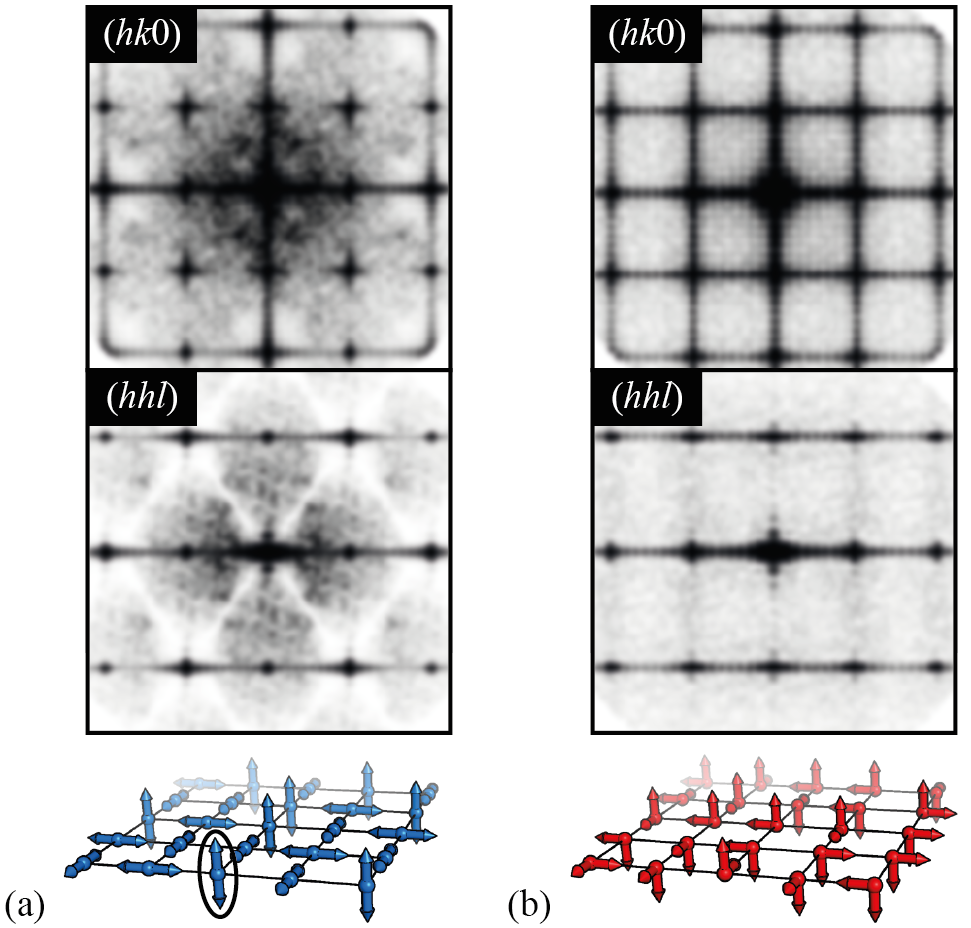}
\end{center}
\caption{\label{fig8}Calculated single-crystal diffraction patterns for (a) 3SP-type and (b) $L$-type orbital disorder models as described in the text. The reciprocal space labels are given relative to the reciprocal lattice of the cubic aristotype. The corresponding orbital-driven distortions are represented figuratively in the configurational sections shown at the bottom of each panel. Our analysis identifies the 3SP-type model as the most likely description of the orbital disordered state; we note that this arrangement destroys the inversion centre at Mn sites (\emph{e.g.}\ for the circled Mn centre).}
\end{figure}

Using a Monte Carlo approach we generated atomistic configurations representing a $20\times20\times20$ supercell of the aristotypic perovskite lattice, in which atomic displacements had been introduced to capture the key correlations in the 3SP- and $L$-type models of orbital disorder in LaMnO$_3$. High-symmetry planes of the corresponding diffraction patterns are shown in Fig.~\ref{fig8}, from which it is evident that---in the absence of thermal disorder---both models would result in highly-structured diffuse scattering as anticipated for correlated disordered states.\cite{Goodwin_2015} In both cases this scattering takes the form of diffuse rods of intensity oriented parallel to the $\langle100\rangle^\ast$ axes of the parent reciprocal lattice, with subtly different reflection conditions within the $(hk0)$ plane. The 3SP model also gives rise to a more structured background scattering pattern, which is most clearly seen here in the $(hhl)$ scattering plane [Fig.~\ref{fig8}(a)].

Despite the existence of some differences between these predicted single-crystal diffuse scattering patterns, there are three reasons why experimental diffuse scattering measurements may nevertheless struggle to conclusively discriminate between the two models. First, our Monte Carlo models are defect-free in the sense that the correlated displacements we include exactly satisfy the constraints of the 3SP- and $L$-type interactions; at the elevated temperatures for which the orbital disorder phase is stable it is reasonable that any such constraints are sometimes broken, resulting in broadening of the corresponding diffuse scattering. So, for example, the structured background expected for the $(hhl)$ plane of the 3SP-type orbital disorder model would almost certainly become washed-out in practice; likewise blurring of the diffuse scattering in the $(hk0)$ plane of the $L$-type orbital disorder model may give the appearance of systematic absences as anticipated for the 3SP-type model. Second, our calculations have been designed to amplify the diffuse scattering (so as to reveal its underlying structure); in reality the intensity would be very low relative to that of the Bragg scattering. In the 3SP-type orbital disorder model, for example, the diffuse scattering arises only from small displacements of O atoms and so in the presence of strong scattering from La would be essentially invisible in X-ray scattering measurements. And, third, the significant degree of thermal motion present within the $O$ phase will degrade the diffraction pattern at high-$Q$ where the differences between in diffuse scattering are clearest. The calculations in Fig.~\ref{fig8} neglect any dynamic contribution to disorder in LaMnO$_3$.

\section{Concluding Remarks}

Our study has cautiously identified that the three state Potts model of Ref.~\citenum{Ahmed_2009} provides the best description of orbital disorder in LaMnO$_3$ in terms of its ability to account for the experimental neutron and X-ray PDFs. Implicit in this model is an isotropic Mn orbital arrangement, which is consistent with the experimental observations of electronic and magnetic isotropy.\cite{Jonker_1956,Goodenough_1999} A more subtle corollary of this model is weak inversion-symmetry breaking at the Mn site [Fig.~\ref{fig8}(a)]. In principle this allows for increased $d$--$p$ mixing, and as such is consistent with the small but reproducible XANES anomalies noted in Refs.~\citenum{Sanchez_2003} to occur at $T_{\textrm{JT}}$. A variation in the local symmetry of Mn environments is also consistent with renormalisation of the Mn-weighted phonon density of states, as suggested by inelastic neutron scattering measurements.\cite{Wdowik_2012} Given the configurational entropy of the 3SP arrangement, one expects low-frequency orbital rearrangements between equivalent 3SP states occurring over a timescale significantly longer than phonon excitations; this is consistent with the NMR treatment of Ref.~\citenum{Trokiner_2013}. Ahmed and Gehring have already established a link\cite{Ahmed_2009} between phase changes in the 3SP model and volume collapse.\cite{Chatterji_2003} Moreover, the entropy calculations of Ref.~\citenum{Wang_1990}---which at face value seemed to establish an inconsistency between the 3SP model and experimental specific heat measurements\cite{Sanchez_2003}---were later revised in Ref.~\citenum{Ahmed_2006} and so the model appears to be consistent with a broad range of experimental observations.

Our experimental PDF data also make clear that the $O^\prime/O$ transition in LaMnO$_3$ is fundamentally different to conventional order--disorder transitions in that there is a discontinuity in the evolution of the PDF at $T_{\textrm{JT}}$. This same point had effectively been noted in the EXAFS and NMR studies of Refs.~\citenum{Souza_2004,Trokiner_2013} but was less clear in earlier PDF studies because of the (understandable) focus on the evolution of the lowest-$r$ Mn--O peak as a function of temperature. From a scientific perspective, the key implication of this discontinuity is that the electronic description of orbital disordered states need not necessarily follow from our understanding of the corresponding ordered states, since orbital arrangements may differ meaningfully between the two. This poses substantial computational challenges because (i) explicit description of disordered orbital arrangements requires large atomistic configurations, and (ii) these states are entropically stabilised and so cannot necessarily be studied meaningfully in the athermal limit. Consequently, we anticipate that further detailed investigation of the orbital disordered states in the broader La$_{1-x}$Ca$_x$MnO$_3$ family---and in particular in the vicinity of the CMR transition---may provide useful insight into CMR itself.

We conclude with a brief discussion of the important methodological limitations of PDF analysis that our study has brought to light. That our preliminary RMC refinements favoured the most disordered description of orbital disorder is hardly surprising: on the one hand, the $L$-type arrangement is simply more likely to be encountered during refinement than either the 3SP- or $C$-type models; and, on the other hand, the difference in quality of fit for the various models is insufficiently large to bias against the statistical result. But what our study emphasises is that ``small-box'' modelling is not itself immune to the uniqueness problem often highlighted only for ``big-box'' approaches (\emph{e.g.}\ RMC, EPSR).\cite{Tucker_2007,Soper_1996} In particular, meaningfully different small-box models give remarkably similar fits to PDF data collected within the orbital disorder regime; even the RMC-derived $L$-type model gives a fit-to-data that is essentially indistinguishable from the published PDF fits of earlier studies [Fig.~\ref{fig6}(b)].\cite{Billinge_2005,Sartbaeva_2007} Hence it may not always be sufficient to present a satisfactory---or even excellent---PDF fit as evidence in support of a particular small-box model. This specific point is of relevance to the problem of disorder in the spin glass Y$_2$Mo$_2$O$_7$, where very different real-space models again give almost-equivalent PDF fits.\cite{Greedan_2009,Thygesen_2017} Here the issue of orbital disorder is resolvable only because the temperatures involved ($T<50$\,K) are so much lower than those of relevance to LaMnO$_3$. Looking forward, one particular challenge for the PDF community is the development of a more complete understanding of which problems are definitively solvable using PDF analysis and which are not. Recent developments such as (i) the 3D-$\Delta$PDF approach,\cite{Simonov_2012} (ii)  ``dynamic'' PDF measurements which can in principle separate static and vibrational contributions to $G(r)$ peak broadening,\cite{Egami_2012,Dmowski_2008} and/or (iii) the ability to incorporate additional data from a variety of experimental techniques during PDF analysis\cite{Levin_2007,Cliffe_2010,Billinge_2010,Juhas_2015} offer a particular sense of optimism in this regard.



\acknowledgments
We acknowledge support in the form of beamtime provision from the ISIS (GEM beamline) and Diamond (I12 beamline; experiment number EE7904) facilities. P.M.M.T., C.A.Y., and A.L.G. acknowledge financial support from the E.P.S.R.C.\ (EP/G004528/2), and the E.R.C. (Grant Ref: 279705).

\bibliography{bibl_LMO}

\end{document}